\pdfoutput=1
\documentclass[12pt, oneside]{article}   	

\usepackage[svgnames]{xcolor}
\usepackage[colorlinks]{hyperref}

\usepackage{geometry}                		
\usepackage{graphicx}				

\usepackage{amssymb}
\usepackage{bm}
\usepackage{amsmath}
\usepackage{bbm}
\usepackage{amsthm}	
\usepackage{natbib}
\usepackage{tikz}
\usepackage{capt-of}
\usepackage{extarrows}
\usepackage{enumitem}  
\usepackage{multirow}
\usepackage{booktabs} 
\usepackage{caption}
\hypersetup{
	citecolor=DarkBlue,
	bookmarksnumbered=true,
	urlcolor=Indigo,linkcolor=blue
}

\setcitestyle{authoryear,open={(},close={)}}

\newtheorem{thm}{Theorem}
\newtheorem{prop}{Proposition}
\newtheorem{ass}{Assumption}

\usepackage{setspace}
\usepackage{csquotes}
 \def\Var{\mathop{\rm Var}} 
\def\LATE{\mathrm{LATE}}
\def\IV{\mathrm{IV}}
\def\AT{\mathrm{AT}}
\def\NT{\mathrm{NT}}
\def\CP{\mathrm{CP}}

\newcommand\independent{\protect\mathpalette{\protect\independenT}{\perp}}
\def\independenT#1#2{\mathrel{\rlap{$#1#2$}\mkern5mu{#1#2}}}

\title{Point Identification of LATE with Two Imperfect Instruments\thanks { I am grateful to Valentina Corradi, Marc Henry, Keisuke Hirano, Sung Jae Jun, Matt Masten,  Joris Pinkse,  Adam Rosen, and Jo\~ao Santos Silva for their many valuable comments.} 
}
\author{Rui Wang%
\thanks{Department of Economics, The Ohio State University.  Email: \texttt{wang.16498@osu.edu}. }
}
\date{December 30, 2022}
\begin{document}
\maketitle
\begin{abstract}

This paper characterizes point identification results of the local average treatment effect (LATE) using two imperfect instruments. The classical approach (\cite{imbens1994}) establishes the identification of LATE via an instrument that satisfies exclusion, monotonicity, and independence. However, it may be challenging to find a single instrument that satisfies all these assumptions simultaneously. My paper uses two instruments but imposes weaker assumptions on both instruments. The first instrument is allowed to violate the exclusion restriction and the second instrument does not need to satisfy monotonicity. Therefore, the first instrument can affect the outcome via both direct effects and a shift in the treatment status. The direct effects can be identified via exogenous variation in the second instrument and therefore the local average treatment effect is identified. An estimator is proposed, and using Monte Carlo simulations, it is shown to perform more robustly than the instrumental variable estimand.

\hspace{10pt} 



\end{abstract}

\newpage

\section{Introduction} \label{sec:intro}
Instrumental variables are widely used to estimate causal effects with endogenous treatment. When treatment effects are heterogeneous, \cite{imbens1994} and \cite*{angrist1996} show that the local average treatment effect (LATE) is identified as the instrumental variable (IV) estimand with a valid instrument. The instrumental variable is required to exert no direct effects on the outcome (exclusion), weakly increase treatment status (monotonicity), and be independent of the potential outcome and potential treatment (independence).

In practice, it is a challenging task to find a valid instrument that satisfies all the assumptions simultaneously. In estimating returns to schooling, several papers such as
\cite{uusitalo1999} use family background variables as an instrument; however, these variables may have direct effects on people's earnings via family education  and thus violate the exclusion restriction. 
The exclusion assumption may also fail in encouragement experiments. For example, \cite{hirano2000} study the effects of the flu vaccine on  the prevalence of influenza and use random encouragement to take the vaccine as an instrument. Their paper  shows that the encouragement to take the vaccine has direct effects on the outcome since it may remind people to take other actions to prevent the flu. 


Studies such as \cite{kitagawa2015}, \cite{huber2015}, \cite{kedagni2016}, and \cite{mourifie2017} develop different methods to test the assumptions of the instrumental variable and reject the validity of some instruments. For example, college proximity is used as an instrument in \cite{card1993}, \cite{kling2001}, and \cite*{carneiro2011}, but its validity is rejected by \cite{huber2015} and \cite{mourifie2017}.

Motivated by these findings, this paper proposes a new approach to identify LATE. This approach uses two instruments while imposing weaker assumptions on both instruments compared to the standard IV assumptions. The first instrument is allowed to violate the exclusion restriction, and the second instrument does not need to satisfy monotonicity. Therefore, the first instrument can affect the outcome through both direct effects and treatment effects.
The second instrument is introduced to separate the two effects. By exploiting exogenous variation in the second instrument, the direct effects of the first instrument are identified, and therefore LATE defined by the first instrument is point identified.

The paper develops estimators for LATE and the direct effects, and establishes their asymptotic properties. I compare this approach by using two imperfect instruments with the IV estimand using a single instrument via Monte Carlo simulations. The results show that the IV estimand can have a large bias with nonzero direct effects and that the bias increases when the direct effects increase.
The method with two instruments presented in this paper performs uniformly regardless of the direct effects; thus, it has a more robust performance concerning violations of the exclusion restriction. 

Here I discuss some potential choices of the two instruments in various applications. One example is the effect of participation in the food stamp program on health outcomes. \footnote{See, e.g., \cite*{debono2012}, \cite{kreider2012}, and \cite*{gundersen2017}.} The first instrument can be an increase in the benefits of the program. Such an increase is likely to affect participation monotonically, but the increased benefit may directly affect health outcomes. The second instrument can be whether the benefit of the program is issued electronically or in hard copy. The delivery method is unlikely to have a direct effect on health outcomes, but it may not satisfy monotonicity,  as some people prefer electronic delivery but others prefer hard-copy delivery.


The second example is the study of the effect of having a third child on mothers' labor supply, as studied in \cite*{angrist1998}. The first instrument can be whether there is a financial subsidy for having a third child. Financial support is likely to encourage a third child, but may directly affect people's incentive for labor participation. The second instrument could be whether the first two children have the same sex. This instrument seems unlikely to directly affect labor participation, but it may violate monotonicity since different people may prefer different sibling-sex composition.


Another example is estimating the effect of early achievement for children (e.g., kindergarten performance) on subsequent outcomes (e.g., earnings), which has been  studied by \cite{chetty2011}. One potential instrument is the random assignment of teachers/classrooms. But as argued by \cite{kolesar2015}, the exclusion restriction might be violated because an experienced teacher may not only improve kindergarten performance, but also have direct effects on children's subsequent outcomes. Nevertheless, we can use a second instrument to separate the direct effects. One potential choice is the class size, which is shown to have no significant effects on earnings in \cite{chetty2011}, but may not satisfy monotonicity as class size effects could be heterogeneous.\footnote{For example, \cite*{maasoumi2005} show opposite signs of class size effects for students below and above the median.}

\subsection{Related Literature}
This paper contributes to the literature studying heterogeneous treatment effects with endogeneity using instrumental variables.
 \cite{imbens1994}, \cite*{angrist1996}, and \cite{heckman2005} identify treatment effects using one instrument that satisfies the conditions of exclusion, monotonicity, and independence simultaneously. My paper complements the literature by using two instruments while relaxing one assumption on each instrument. 


Several papers relax the exclusion restriction under the heterogeneous treatment effects framework. \cite{hirano2000} study different violations of exclusion restrictions for subgroups and apply parametric models and the Bayesian approach for inference.
\cite{flores2013} derive partial identification for LATE by employing a weak monotonicity assumption 
of mean potential outcomes within or across subgroups. \cite{mealli2013} study partial identification of  intention-to-treat effects through a secondary outcome. 


\cite{de2017} relaxes the monotonicity assumption and shows that the IV estimand estimates the local average treatment for a subgroup of compliers under a ``compliers-defiers'' condition. He further provides sufficient conditions for the ``compliers-defiers'' condition to hold. \cite{kedagni2021} relaxes the independence assumption of the instrument with potential treatment. His paper establishes partial identification results by using an additional instrument that serves as a proxy of the first instrument.


The paper is also related to the literature that relaxes the exclusion restriction of the instrument in a linear regression model. \cite{hahn2005} derive the bias of different estimators with direct effects. \cite{nevo2012} provide partial identification for the model parameter under the assumptions of a correlation between the instrument and the error term.  \cite*{conley2012} employ different assumptions on the effect of the instrument on the outcome to conduct inference for the model parameter. 
\cite{kolesar2015}  allow for direct effects and develop an estimator under an orthogonality condition of the direct effects.

The remainder of the paper is organized as follows. Section \ref{sec:treat} presents the heterogeneous treatment effect framework. Section \ref{sec:iden} derives the identification result. Section \ref{sec:esti} develops an estimator for LATE and examines its finite sample performance via simulations. Section \ref{sec:exte} studies an extension and Section \ref{sec:conc} concludes.

\section{Heterogeneous Treatment Effects Model} \label{sec:treat}
The analysis focuses on the heterogeneous treatment effects framework introduced in \cite{imbens1994} and  \cite*{angrist1996}. Let $Y\in \mathcal{Y}$ denote an outcome, $D \in\{0 ,1\}$ denote a binary treatment, and $Z\in \{0, 1\}$ denote a binary instrument. The objective is to learn the effects of treatment $D$ on outcome $Y$, but the treatment can be endogenous. Instrument $Z$ is used to address the endogeneity issue. Observed variables are $(Y, D, Z)$.
 
Following \cite{imbens1994} and \cite*{angrist1996}, I use counterfactual variables to describe the data generating process. Let $D_z$ denote the potential treatment given the instrument $Z=z$ and $Y_{d, z}$ denote the potential outcome given the instrument and treatment $Z=z, D_z=d$. Let $Y_d$ denote the potential outcome given $D=d$, which is given as $Y_d=Y_{d, 1}Z+Y_{d, 0}(1-Z)$. Observed variables $(Y, D)$ are generated by
\begin{equation*}
\begin{aligned}
D&=D_1Z+D_0(1-Z), \\
Y&=Y_1 D+Y_0 (1-D).
\end{aligned}
\end{equation*}

The population can be divided into four subgroups based on the potential treatments $(D_1, D_0)$: always takers (AT): $D_1=D_0=1$; compliers (CP): $D_z=z$ for $z\in \{0, 1\}$; never takers (NT):  $D_1=D_0=0$; and defiers (DF): $D_z=1-z$ for $z\in \{0, 1\}$.

The following summarizes the assumptions in \cite{imbens1994} and \cite*{angrist1996}.

\begin{ass}[IV Validity] \label{ass:IV}

\
%
%
%
%
%
%
%
\begin{enumerate}[label=(\roman*)]

\item Exclusion: $Y_{d,1}=Y_{d, 0}\equiv Y_d$ for any $d\in \{0, 1\}$;  
\vspace{-0.2cm}
\item Monotonicity: $D_1\geq D_0$;
\vspace{-0.2cm}
\item Independence: $Z\independent (Y_{1,1}, Y_{1,0}, Y_{0, 1}, Y_{0, 0}, D_1, D_0)$;
\vspace{-0.2cm}
\item Relevance: $\Pr(D_1>D_0)>0$; 
\vspace{-0.2cm}
\item $0<\Pr(Z=1)<1$.

\end{enumerate}

\end{ass}

Assumption \ref{ass:IV} (i) requires instrument $Z$ to have no direct effects on the potential outcome, (ii) indicates that the instrument weakly increases the potential treatment (no defier), (iii) refers to the independence of the instrument with all potential variables, (iv) guarantees the existence of compliers, and (v) needs nonzero variation in instrument $Z$.

Under Assumption \ref{ass:IV}, \cite{imbens1994} show that the average treatment effect for compliers can be identified as the IV estimand:
\begin{equation}\label{equ:late}
\begin{aligned}
\LATE\equiv E[Y_1-Y_0\mid \CP]=\frac{E[Y\mid Z=1]-E[Y\mid Z=0]}{E[D\mid Z=1]-E[D\mid Z=0]}\equiv \IV.
\end{aligned}
\end{equation}

This identification result for LATE relies on the validity of instrument $Z$ in Assumption \ref{ass:IV}. However, it may be difficult to find a single instrument satisfying all conditions in Assumption \ref{ass:IV}. The identification result may fail if one of the assumptions does not hold. 

 \cite*{angrist1996} (Section 5) discuss the sensitivity of the IV estimand to deviations from the IV validity assumptions. The IV estimand involves both direct effects and treatment effects when the exclusion restriction fails, and it is the combination of treatment effects for compliers and defiers when the monotonicity condition is violated. This paper proposes a new approach to identify LATE by using two instruments jointly but requiring weaker assumptions on both instruments. The first instrument is allowed to violate the exclusion restriction, and the second instrument does not need to satisfy the monotonicity condition.

\section{Identifying LATE with Two Invalid Instruments} \label{sec:iden}

This section focuses on identifying the local average treatment effect defined by instrument $Z$, while allowing instrument $Z$ to have nonzero direct effects on the outcome. I introduce an additional binary instrument $W\in \{0, 1\}$ to identify the direct effects and LATE defined by instrument $Z$. 
My approach uses two instruments $(Z, W)$ but relaxing one main assumption on both instruments. Instrument $Z$ is allowed to have direct effects on the outcome, and instrument $W$ does not need to satisfy monotonicity.  Instrument $W$ is assumed to have no direct effects on the outcome such that the potential outcome is not indexed by instrument $W$.

When the exclusion restriction is relaxed, instrument $Z$ will affect the outcome through both direct effects and treatment effects. The direct effect is $Y_{d, 1}-Y_{d, 0}$ for each individual with $D=d$. Let $\rho_{G, d}=E[Y_{d,1}-Y_{d, 0}\mid G]$ denote the average direct effect for group $G\in\{\AT, \NT, \CP\}$ given treatment $D=d$. 

Next, I present assumptions on the two instruments $(Z, W)$.

\begin{ass}[Instruments $Z \& W$] \label{ass:z} 
\

\begin{enumerate}[label=(\roman*)]
\item Direct effects: $\rho_{G, d}=\rho_d$ for any $G\in \{\AT, \NT, \CP\}$, where $\rho_d$ is a unknown constant for $d\in \{1, 0\}$; 
\vspace{-0.2cm}
\item Monotonicity: $D_1\geq D_0$;  
\vspace{-0.2cm}
\item Independence: $(Z, W)\independent Y_{d, z} \mid (D_1, D_0)$ for any $d, z\in \{0, 1\}$, and $Z\independent (D_1, D_0) \mid W$;
\vspace{-0.2cm}
\item Relevance: $\Pr(D_1>D_0)>0$; 
\vspace{-0.2cm}
\item $0<\Pr(Z=z, W=w)<1$ for any $z, w\in\{0, 1\}$.

\end{enumerate}
\end{ass}

Assumption \ref{ass:z} (i) relaxes the exclusion restriction in Assumption \ref{ass:IV} and allows instrument $Z$ to have nonzero direct effects on the outcome, $\rho_d\neq 0$. The average of the direct effects is assumed to be homogeneous across different subgroups but may vary given different treatments, $\rho_1\neq \rho_0$. One example for potential outcome $Y_{d, z}$ is given as $Y_{d, z}=\rho_d z+u_d$, where $\rho_d$ is a unknown constant. Then the direct effect is $\rho_d$ for each individual given treatment $D=d$. Section \ref{sec:exte} studies an extension allowing for different direct effects across subgroups and provides partial identification of LATE. The rest of the conditions in Assumption \ref{ass:z} are similar to Assumption \ref{ass:IV} except for introducing the independence condition of instrument $W$ with the potential outcome.

Under the monotonicity condition of instrument $Z$, we can divide the population into three subgroups $\{\AT, \NT, \CP\}$ as described in Section \ref{sec:treat}. The objective is to identify the average treatment effect for the compliers defined by instrument $Z$:
\begin{equation*}
\LATE=E[Y_1-Y_0\mid \CP].
\end{equation*}

Under the assumption of no direct effects ($\rho_1=\rho_0=0$), LATE is identified as
\begin{equation*}
\LATE=\IV_1\equiv \frac{E[Y\mid Z=1, W=1]-E[Y\mid Z=0, W=1]}{E[D\mid Z=1, W=1]-E[D\mid Z=0, W=1]}.
\end{equation*}

The $\IV_1$ estimand is similar to the $\IV$ estimand except it is conditional on the additional variable $W$. However, the above identification result does not apply if $Z$ has nonzero direct effects on the outcome. 
My paper uses the additional instrument $W\in \{0, 1\}$ to help identify the direct effects $(\rho_1, \rho_0)$ and LATE.  

The following describes a relevance condition of instrument $W$.

\begin{ass}[Instrument W]\label{ass:w}
Relevance: $\Pr(G \mid W=1)\Pr(\CP\mid W=0)\neq \Pr(G \mid W=0)\Pr(\CP\mid W=1)$ for any $G \in \{\AT, \NT\}$.
\end{ass}

Assumption \ref{ass:w} is a relevance condition of instrument $W$. It requires that instrument $W$ is correlated with the potential treatment such that the size of always takers (or never takers) relative to compliers varies when instrument $W$ changes. Assumption \ref{ass:w} does not impose monotonicity on instrument $W$, so instrument $W$ can affect treatment in any direction. This assumption is testable since the conditional probability of all types $G\in \{\AT, \NT, \CP\}$ is identified, as shown in Appendix \ref{proof:thm1}. 


Section \ref{sec:intro} provides some examples of the two instruments in various applications. When estimating the effects of the food stamp program on health outcomes, the first instrument can be an increase in the benefits of the program. The increased benefits may directly affect health outcomes so the exclusion restriction fails. The second instrument can be whether the benefit of the program is issued electronically or in hard copy.
This instrument may not satisfy the monotonicity condition since some people prefer electronic delivery but others prefer hard-copy delivery.

\begin{thm} \label{thm:point}
Under Assumptions \ref{ass:z}-\ref{ass:w},  direct effects $(\rho_1, \rho_0)$ and local average treatment effect $\LATE$ are point identified.
\end{thm}

When the exclusion restriction is relaxed, instrument $Z$ can induce both treatment effects by switching the treatment status and direct effects on the outcome. We are unable to distinguish treatment effects and direct effects without further information, so the standard identification result for treatment effects no longer applies. This paper uses additional instrument $W$, which can serve as an instrument for the imperfect instrument $Z$, to address instrument $Z$'s violation of the exclusion restriction. By exploiting variation in instrument $W$, this approach can identify the direct effects of instrument $Z$ on outcome $Y$. Then the local average treatment effect is identified by subtracting the direct effects.

Theorem \ref{thm:point} provides an alternative approach to identify treatment effects. This approach uses two instruments while relaxing one of the assumptions on the two instruments. This method can be applied to scenarios where there are multiple options of instruments but the instruments are imperfect in different dimensions. The availability of multiple instruments is discussed in the literature including \cite{card2001}, \cite{hausman2012}, and \cite{kolesar2015}. Moreover, instrument $W$ can help identify the direct effects of instrument $Z$ on the treated and untreated outcome, meaning that this approach can be used to test whether instrument $Z$ has direct effects on the outcome. 

%

\section{Estimation} \label{sec:esti}
This section provides an estimation method for direct effects $(\rho_1, \rho_0)$ and local average treatment effect $\LATE$. Suppose that we have an $i.i.d$ sample $(Y_i, D_i, Z_i, W_i)_{i=1}^N$. 
As shown in Appendix \ref{proof:thm1}, LATE is identified as the $\IV_1$ estimand minus a weighted average of the two direct effects $(\rho_1, \rho_0)$:
\begin{equation*}
\LATE=\IV_1-\rho_1w^{\rho}_{1}-\rho_0 w^{\rho}_0,
\end{equation*}
where $w^{\rho}_1=\frac{\Pr(\AT\mid W=1)}{\Pr(\CP\mid W=1)}+\Pr(Z=0)$, $w^{\rho}_0=\frac{\Pr(\NT\mid W=1)}{\Pr(\CP\mid W=1)}+\Pr(Z=1)$, and the formula for the two direct effects $(\rho_1, \rho_0)$ is described below. 

To estimate $\LATE$, we need to estimate all terms in the above expression of $\LATE$. The first term $\IV_1$ is given as
\begin{equation*}
\begin{aligned}
\IV_1&=\frac{E[Y\mid Z=1, W=1]-E[Y\mid Z=0, W=1]}{E[D\mid Z=1, W=1]-E[D\mid Z=0, W=1]} \\
&=\frac{E[YZW]E[W]-E[YW] E[ZW]}{E[DZW] E[W]-E[DW]E[ZW]}.
\end{aligned}
\end{equation*}

The estimator for the $\IV_1$ estimand can be developed by replacing the population expectation with the sample mean:
\begin{equation*}
\widehat{\IV}_1=\frac{\sum_i(Y_iZ_iW_i) \sum W_i-\sum_i(Y_iW_i)\sum_i(Z_iW_i)}{\sum_i(D_iZ_iW_i)  \sum W_i-\sum_i(D_iW_i)\sum_i(Z_iW_i)}.
\end{equation*}

Now I construct estimators for the two weights $(w_1^{\rho}, w_0^{\rho})$. 
Let $\widehat{\Pr}(G\mid w)$ denote the estimator for the conditional probability $\Pr(G\mid w)$ of the three subgroups $G\in\{\AT, \NT, \CP\}$ given $W=w$, constructed as follows: 
\begin{equation*}
\begin{aligned}
\widehat{\Pr}(\AT\mid w)&=\frac{\sum_i D_i(1-Z_i)\mathbbm{1}\{W_i=w\} }{\sum_i (1-Z_i)\mathbbm{1}\{W_i=w\}}, \\
\widehat{\Pr}(\NT\mid w)&=\frac{\sum_i (1-D_i)Z_i \mathbbm{1}\{W_i=w\} } {\sum_i Z_i\mathbbm{1}\{W_i=w\}}, \\
\widehat{\Pr}(\CP\mid w)&=\frac{\sum_i D_iZ_i\mathbbm{1}\{W_i=w\} }{\sum_i Z_i\mathbbm{1}\{W_i=w\}}-\frac{\sum_i D_i(1-Z_i)\mathbbm{1}\{W_i=w\} }{\sum_i (1-Z_i)\mathbbm{1}\{W_i=w\}}.
\end{aligned}
\end{equation*}


Then the two weights $(w^{\rho}_1, w^{\rho}_0)$ can be estimated as
\begin{equation*}
\begin{aligned}
\hat{w}^{\rho}_1&=\frac{\widehat{\Pr}(\AT \mid 1)}{\widehat{\Pr}(\CP\mid 1)}+1-\bar{Z}, \qquad
\hat{w}^{\rho}_0&=\frac{\widehat{\Pr}(\NT\mid 1)}{\widehat{\Pr}(\CP\mid 1)}+\bar{Z},
\end{aligned}
\end{equation*}
where $\bar{Z}=\frac{1}{N}\sum_i Z_i$.

We only need to develop estimators for the two direct effects $(\rho_1, \rho_0)$. I focus on the estimator for direct effect $\rho_1$, and the idea also applies to $\rho_0$. As shown in Appendix \ref{proof:thm1}, direct effect $\rho_1$ is identified as
\begin{equation*}
\rho_1=\frac{r_1(1)\Pr(\CP\mid 0)-r_1(0)\Pr(\CP\mid 1) }{\Pr(\AT\mid 1)\Pr(\CP\mid 0)-\Pr(\AT\mid 0)\Pr(\CP\mid 1) },
\end{equation*}
where $r_1(w)$ is defined as
\begin{equation*}
r_1(w)\equiv E[YD\mid Z=1, w]-E[YD\mid Z=0, w].
\end{equation*}

The estimator for $r_1(w)$ is developed as follows:
\begin{equation*}
\hat{r}_1(w)=\frac{\sum_i Y_iD_iZ_i\mathbbm{1}\{W_i=w\}}{\sum_i Z_i\mathbbm{1}\{W_i=w\}}-\frac{\sum_i Y_iD_i(1-Z_i)\mathbbm{1}\{W_i=w\}}{\sum_i (1-Z_i)\mathbbm{1}\{W_i=w\}}.
\end{equation*}

Then estimator $\hat{\rho}_1$ for direct effect $\rho_1$ can be established by replacing all terms with their estimators:
\begin{equation*}
\hat{\rho}_1=\frac{\hat{r}_1(1)\widehat{\Pr}(\CP\mid 0)-\hat{r}_1(0)\widehat{\Pr}(\CP\mid 1) }{\widehat{\Pr}(\AT\mid 1)\widehat{\Pr}(\CP\mid 0)-\widehat{\Pr}(\AT\mid 0)\widehat{\Pr}(\CP\mid 1) }. 
\end{equation*}

Estimator $\hat{\rho}_0$ for $\rho_0$ can be established similarly, so it is omitted here. 
Local average treatment effect $\LATE$ is estimated as
\begin{equation*}
\widehat{\LATE}=\widehat{\IV}_1-\hat{\rho}_1 \hat{w}^{\rho}_1-\hat{\rho}_0 \hat{w}^{\rho}_0.
\end{equation*}

The asymptotic properties of $\widehat{\LATE}$ and the two direct effects $(\hat{\rho}_1, \hat{\rho}_0)$ are derived in Appendix \ref{asymptotic}. The estimators $\widehat{\LATE}$ and $(\hat{\rho}_1, \hat{\rho}_0)$ are $\sqrt{N}$ consistent assuming all variances and covariances are finite, and the expressions of asymptotic variances are shown in Appendix \ref{asymptotic}. 

 \subsection{Simulation Study}
 
This section examines the finite sample performance of estimator $\widehat{\LATE}$ by using two instruments $(Z, W)$ via Monte Carlo simulations. I compare the estimator in this paper with the IV estimator in \cite{imbens1994} and show that the method in this paper works more robustly when the direct effects are nonzero. The $\IV_1$ estimator (conditional on $W$) performs slightly worse than the $\IV$ estimator, so I display the results of the $\IV$ estimator for comparison.

The simulation setup is as follows. Instrument $Z$ follows the Bernoulli distribution with probability $p=0.5$. Potential treatment $D_z$ is $D_z=\mathbbm{1}\{z\geq \epsilon \}$, and observed treatment $D$ is $D=D_1 Z+D_0(1-Z)$.
Potential outcome $Y_d$ is given $Y_d=a_d+\rho_d Z+u_d$ for $d\in\{0, 1\}$, where $a_1=1$, $a_0=0$. Observed outcome $Y$ is given $Y=Y_1D+Y_0(1-D)$. I consider four different cases of direct effects: $\rho_1=\rho_0=\rho\in \{0, 0.5, 1, -1\}$. 
Instrument $W$ is given as $W=\mathbbm{1}\{v\leq D_1+D_0\}$. 

Latent variables $(\epsilon, u_1)$ follow a standard multivariate distribution with correlation $c=0.5$, and this correlation captures the endogeneity of treatment $D$. Error terms $u_0$ and $v$ follow a standard normal distribution and are independent of all other variables. I consider sample size $N=\{1000, 4000, 16000\}$, and the repetition number is $B=5000$.

The assumptions of the two instruments $(Z, W)$ are satisfied under this setup. Instrument $Z$ is allowed to have nonzero direct effects on outcome $\rho_d\neq 0$, and the direct effects are the same across subgroups. The independence condition of $Z$ is satisfied since it is independent of all variables. The monotonicity and relevance conditions are satisfied by the definition of potential treatment $D_z$. Instrument $W$ is independent of the potential outcome given the potential treatment since error term $v$ is independent of all variables. The relevance of instrument $W$ is satisfied because $W$ depends on the potential outcome.

Let $\hat{\theta}_{zw}$ denote the estimator by using the two instruments $(Z, W)$ in this paper, and let $\hat{\theta}_z$ denote the IV estimator by using only instrument $Z$ in \cite{imbens1994}. Let $(\hat{\rho}_1, \hat{\rho}_0)$ denote the estimators for the direct effects $(\rho_1, \rho_0)$ in this paper. To compare the two estimators $(\hat{\theta}_{zw}, \hat{\theta}_{z})$, I report four evaluations of the two estimators: bias, standard deviation (SD), root mean-squared error (rMSE), and median of absolute deviation (MAD). Under the simulation setup, the true local average treatment effect $\theta_0$ is given as
\begin{equation*}
\begin{aligned}
\theta_0=E[Y_1-Y_0\mid \CP]=&E[a_1-a_0+(\rho_1-\rho_0)Z+u_1-u_0\mid \epsilon\in(0, 1)] \\
=&1+\frac{(\phi(0)-\phi(1))}{2(\Phi(1)-\Phi(0))},
\end{aligned}
\end{equation*}
where $\phi$ and $ \Phi$ denote the PDF and CDF of the standard normal distribution, respectively.

Table \ref{table:est} presents the performance of the two estimators $\hat{\theta}_{zw}$ and $\hat{\theta}_z$ under different direct effects $\rho\in \{0, 0.5, 1, -1\}$ and sample size $N\in\{1000, 4000, 16000\}$. Estimator $\hat{\theta}_z$ performs better when the direct effect is zero but can have a large bias with nonzero direct effects. The bias increases when the direct effect is larger, and it does not decrease even when the sample size increases. When the direct effect is negative $\rho<0$, the bias becomes negative and estimator $\hat{\theta}_z$ may have the wrong signs of the true treatment effect. 
Estimator $\hat{\theta}_{zw}$ by using two instruments performs uniformly under different direct effects, and it performs better than estimator $\hat{\theta}_z$ with nonzero direct effects. This pattern becomes more significant when the sample size increases. 
The comparison in Table \ref {table:est} shows that estimator $\hat{\theta}_{zw}$ by using two instruments has a more robust performance for nonzero direct effects.

\begin{table}[!htbp]
\centering
\caption{Performance Comparisons of $\hat{\theta}_{zw}$ and $\hat{\theta}_z$ }
\label{table:est}
\begin{tabular}{cc |cccc|cccc}
\hline
\hline
 \multirow{2}{*}{$N$}&\multirow{2}{*}{$\rho$}&\multicolumn{4}{c|}{$\hat{\theta}_{zw}$}  &  \multicolumn{4}{c}{$\hat{\theta}_z$}  \\
\cline{3-10} 
&  & Bias & SD & rMSE & MAD  & Bias & SD  & rMSE & MAD \\
\hline
\multirow{5}{*}{1000} & $0$ \ & 0.025 & 0.504 & 0.505 & 0.396 & 0.006 &0.183 &0.184 &0.146  \\ [1.0ex]       
& $0.5$ \ & 0.025 & 0.504 & 0.505 & 0.396 &1.481 & 0.245 & 1.501 &1.481  \\ [1.0ex]                                                                                                                                                                                                                                                                                                                                                                                                                                                                                                                                                                                                                                                                    
&  $1$ \  & 0.025 & 0.504 & 0.505 & 0.396  & 2.956 & 0.342 & 2.976 & 2.956 \\ [1.0ex]  
& $-1$ \  & 0.025 & 0.504 & 0.505 & 0.396  & -2.944 & 0.269 &2.956 &2.944  \\ [1.0ex] 
\hline
\multirow{5}{*}{4000} & $0$ \ &0.009 &0.230 &0.230 & 0.183 &-0.001 &0.092 &0.092 &0.073  \\ [1.0ex] 
&  $0.5$ \  &0.009 &0.230 &0.230 & 0.183 & 1.466 &0.119 & 1.470 & 1.466   \\ [1.0ex] 
& $1$ \  &0.009 &0.230 &0.230 & 0.183&  2.932 & 0.165 &2.936 &2.932      \\ [1.0ex] 
& $-1$ \  &0.009 &0.230 &0.230 & 0.183 & -2.933 &0.133 &2.936 &2.933     \\   [1.0ex] 
\hline
\multirow{5}{*}{16000} & $0$ \ &0.001 &0.113 &0.113 &0.090 &0.001 &0.046 &0.046 &0.036  \\ [1.0ex]       
&  $0.5$ \  &0.001 &0.113 &0.113 &0.090 & 1.465 & 0.060 &1.466 &1.465  \\ [1.0ex]                                                                                                                                                                                                                                                                                                                                                                                                                                                                                                                                                                                                                                                                 
& $1$ \ &0.001 &0.113 &0.113 &0.090& 2.930 & 0.083 & 2.931  & 2.930 \\ [1.0ex]   
&  $-1$ \ &0.001 &0.113 &0.113 &0.090 &  -2.929 & 0.066 &2.930 &2.929 \\ [1.0ex] 
\hline
\end{tabular}
\end{table}
\newpage
Table \ref{table:direct} presents the performance of the estimators $(\hat{\rho}_1, \hat{\rho}_0)$ for the direct effects. The performance of $(\hat{\rho}_1, \hat{\rho}_0)$ does not depend on the true direct effects, so I only report the results under different sample sizes. When the sample size increases, the bias and deviation of  $(\hat{\rho}_1, \hat{\rho}_0)$ shrink dramatically. 

Appendix \ref{simu} presents more simulation results about the two estimators $(\hat{\theta}_{zw}, \hat{\theta}_z)$  under different probabilities of the three subgroups.

\begin{table}[!htbp]
\centering
\caption{Performance of Direct Effects $\hat{\rho}_1$ and $\hat{\rho}_0$}
\label{table:direct}
\begin{tabular}{c |cccc|cccc}
\hline
\hline
\rule{0pt}{20pt} \multirow{2}{*}{$N$}&\multicolumn{4}{c|}{$\hat{\rho}_1$}  &  \multicolumn{4}{c}{$\hat{\rho}_0$}  \\
\cline{2-9} 
 \rule{0pt}{20pt}& Bias & SD & rMSE & MAD  & Bias & SD  & rMSE & MAD \\
\hline
\rule{0pt}{20pt}1000 & -0.015 & 0.210 & 0.210 & 0.160 & 0.006 &0.351 &0.351 &0.275  \\ [1.7ex]       
\hline
\rule{0pt}{20pt}4000  & -0.004 &0.095 &0.095 &0.075 &-0.003 & 0.168 &0.168 &0.133  \\ [0.7ex] 
\hline
\rule{0pt}{20pt}16000 & 0.000 &0.047 &0.047 &0.038 &-0.001 & 0.083 &0.083 &0.066  \\ [0.7ex]       
\hline
\end{tabular}
\end{table}

\section{Extension} \label{sec:exte}

Section \ref{sec:iden} establishes point identification results of LATE when the average of direct effects is assumed to be homogeneous across subgroups. This assumption can allow for heterogeneous direct effects within a subgroup but assumes same direct effects across subgroups. This section further relaxes this assumption and provides partial identification results under heterogeneous direct effects across subgroups. 

Recall that $\rho_{G, d}=E[Y_{d,1}-Y_{d, 0}\mid G]$ denote the average direct effect for group $G\in\{\AT, \NT, \CP\}$ given treatment $D=d$. I consider that the heterogeneity in direct effects between subgroups can be bounded by a known number.

\begin{ass} \label{ass:dir}
Direct effects: $|\rho_{\CP, 1}-\rho_{\AT, 1}|\leq k_1$ and $|\rho_{\CP, 0}-\rho_{\NT, 0}|\leq k_0$, where $k_1, k_0\geq 0$ are known constants.
\end{ass}

Assumption \ref{ass:dir} relaxes Assumption \ref{ass:z} (i) and allows heterogeneous direct effects across different subgroups. This assumption requires that the difference in direct effects between subgroups is not too large and can be bounded by a known number $k_d$. The information about $k_d$ depends on specific applications. For example, when the instrument is whether there is an increase (or decrease) in the benefits of a social program, the direct effect on people who did not participate in the program is zero $\rho_{\NT, 0}=\rho_{\CP, 0}=0$ so that $k_0=0$. The value of $k_1$ can be developed when the support of the outcome is bounded such as binary outcomes.



As shown in Section \ref{sec:esti}, when the average of direct effects is the same across different groups $k_1=k_0=0$, $\LATE$ is identified as 
\begin{equation*}
\LATE=\IV_1-\rho_1 w^{\rho}_1-\rho_0 w^{\rho}_0\equiv \widetilde{\IV}. 
\end{equation*}

The next proposition derives sharp bounds on $\LATE$.

\begin{prop} \label{prop:exte}
Under Assumptions \ref{ass:z} (ii)-(iv) and Assumptions \ref{ass:w}-\ref{ass:dir}, the sharp bounds for $\LATE$ are given as follows:
\[ \LATE\in \left[\widetilde{\IV}-k_1 \Pr(Z=0)-k_0 \Pr(Z=1),  \widetilde{\IV}+k_1 \Pr(Z=0)+k_0 \Pr(Z=1)  \right]. \]

\end{prop}

Proposition \ref{prop:exte} shows that $\LATE$ can be still bounded by using two instruments under heterogeneous direct effects. The bounds are tighter when the heterogeneity $(k_1, k_0)$ in direct effects between different subgroups is smaller, and point identification is achieved when the heterogeneity is zero. The bounds on LATE can be established similarly and be further tightened if the direction of the difference in direct effects is known such as $0\leq \rho_{\CP, 1}-\rho_{\AT, 1}\leq k_1$ and $0\leq \rho_{\CP, 0}-\rho_{\NT, 0}\leq k_0$.

\section{Conclusion} \label{sec:conc}

This paper proposes a new approach to point identify LATE by using two instruments while imposing weaker assumptions on both instruments compared to \cite{imbens1994}. The first instrument is allowed to violate the exclusion restriction, so it can have nonzero direct effects on the outcome. Then the $\IV$ estimand involves both treatment effects and direct effects, so the standard identification result does not apply. This paper uses an additional imperfect instrument, which does not need to satisfy monotonicity. By exploiting variation in the second instrument, we can identify the direct effects of the first instrument and then identify LATE.
Based on the identification results, an estimator for LATE is developed and it is shown to perform more robustly than the $\IV$ estimator with nonzero direct effects.

This paper relaxes different assumptions for the two instruments to achieve point identification of treatment effects. It would be worthwhile to investigate the identifying power of multiple instruments when they violate the same assumption such as the exclusion restriction. Moreover, the paper considers two instruments and relaxes one main assumption for each of the two instruments. It would be interesting to explore whether point identification can be achieved under weaker conditions on instruments when there are more than two instruments available.

\bibliography{LATE}
\bibliographystyle{apalike}

\appendix

\section{Appendix} \label{sec:appen}

\subsection{Proof of Theorem \ref{thm:point}} \label{proof:thm1}
                  
\begin{proof}

I first look at the expression of LATE, and divide it into two groups $Z=1$ and $Z=0$ as follows:
\begin{equation*}
\begin{aligned}
\LATE=&E[Y_1-Y_0\mid \CP]\\
=&E[Y_{1, 1}-Y_{0, 1}\mid \CP, Z=1]\Pr(Z=1)+E[Y_{1, 0}-Y_{0, 0}\mid \CP, Z=0]\Pr(Z=0)\\
=&E[Y_{1, 1}-Y_{0, 1}\mid \CP]\Pr(Z=1)+E[Y_{1, 0}-Y_{0, 0}\mid \CP]\Pr(Z=0).
\end{aligned}
\end{equation*}
The above condition holds by the independence condition of instrument $Z$ in Assumption \ref{ass:z}. Using the condition $E[Y_{d, 1}-Y_{d, 0}\mid \CP]=\rho_d$ to substitute $E[Y_{0, 1} \mid \CP]$ and $E[Y_{1, 0} \mid \CP]$ leads to the following implication:
\begin{equation*}
\LATE=E[Y_{1, 1}-Y_{0, 0}\mid \CP]-\rho_1\Pr(Z=0)-\rho_0\Pr(Z=1).
\end{equation*}

To prove LATE is identified, we need to show that the two direct effects $(\rho_1, \rho_0)$ and $(E[Y_{11}\mid \CP], E[Y_{0, 0}\mid \CP])$ are identified.  To prove it, I first show that the conditional probability of the three subgroups $\{\AT, \NT, \CP\}$ given $W=w$ is identified. When $Z=0$, only always takers (AT) are treated so that the probability of always takers is identified as
\begin{equation*}
\begin{aligned}
\Pr(D=1\mid Z=0,w)&=\Pr(D=1\mid \AT, Z=0,w)\Pr(\AT\mid Z=0, w)\\
&=\Pr(\AT\mid w). \\
\end{aligned}
\end{equation*}
The last equality holds since the probability of being treated conditional on always takers is one and instrument $Z$ is independent of potential treatments given $W$ in Assumption \ref{ass:z}.

Similarly, the conditional probability of never takers and compliers given $W=w$ can be derived as

\begin{equation*}
\begin{aligned}
\Pr(\NT\mid w)=&\Pr(D=0\mid Z=1,w);   \\
\Pr(\CP\mid w)=&\Pr(D=1\mid Z=1,w)-\Pr(D=1\mid Z=0,w).
\end{aligned}
\end{equation*}


Now we are ready to show that the two direct effects $(\rho_1, \rho_0)$ and $(E[Y_{11}\mid \CP], E[Y_{0, 0}\mid \CP])$ are identified by using variation in instrument $W$.
The expectation of $YD$ conditional on $(Z=1, W=w)$ can be the expressed as a mixture of always takers and compliers:
\begin{equation} \label{z1}
\begin{aligned}
&E[YD\mid Z=1, w]\\
=&E[Y_{1,1}\mid \AT, Z=1, w]\Pr(\AT\mid Z=1, w)+E[Y_{1, 1}\mid \CP, Z=1, w]\Pr(\CP\mid Z=1, w) \\
=&E[Y_{1, 1}\mid \AT]\Pr(\AT\mid w)+E[Y_{1,1}\mid \CP]\Pr(\CP\mid w). 
\end{aligned}
\end{equation}
The above condition holds by the independence conditions of the two instruments $(Z, W)$ in Assumptions \ref{ass:z}.

Similarly, the expectation of $YD$ given $(Z=0, W=w)$ can be expressed as
\begin{equation}\label{z0}
E[YD\mid Z=0, w]=E[Y_{1, 0}\mid \AT]\Pr(\AT\mid w).
\end{equation}

Taking the difference between \eqref{z1} and \eqref{z0} leads to the following condition: for any $w\in\{0, 1\}$,
\begin{equation}\label{eq:dif}
\begin{aligned}
r_1(w)&\equiv E[YD\mid Z=1, w]-E[YD\mid Z=0, w]\\
&=\rho_1\Pr(\AT\mid w)+E[Y_{1,1}\mid \CP]\Pr(\CP\mid w).
\end{aligned}
\end{equation}

Equation \eqref{eq:dif} comes from the condition $E[Y_{d, 1}-Y_{d, 0}\mid \AT]=\rho_d$ in Assumption \ref{ass:z} (i). Since condition \eqref{eq:dif} holds for any $w\in \{0 ,1\}$, using variation in  instrument $W$ can identify direct effect $\rho_1$ and $E[Y_{1, 1}\mid \CP]$ as follows:
\begin{equation*}
\begin{aligned}
&\rho_1=\frac{r_1(1)\Pr(\CP\mid 0)-r_1(0)\Pr(\CP\mid 1) }{\Pr(\AT\mid 1)\Pr(\CP\mid 0)-\Pr(\AT\mid 0)\Pr(\CP\mid 1) }, \\
&E[Y_{1,1}\mid \CP]=\frac{r_1(1)-\rho_1\Pr(\AT\mid 1) }{\Pr(\CP\mid 1 ) }. \\
\end{aligned}
\end{equation*}
The relevance condition of instrument $W$ in Assumption \ref{ass:w} guarantees that the denominator of  direct effect $\rho_1$ is nonzero: $\Pr(\AT\mid 1)\Pr(\CP\mid 0)-\Pr(\AT\mid 0)\Pr(\CP\mid1)\neq 0$.  The relevance condition of instrument $Z$ in Assumption \ref{ass:z} implies that there exists $w\in \{0, 1\}$ such that $\Pr(\CP\mid w)>0$. For simplicity, I assume that $\Pr(\CP\mid 1)>0$.

Similarly, I use the expectation of $Y(1-D)$ under different values of $(Z, W)$ to identify direct effect $\rho_0$ and $E[Y_{0, 0}\mid \CP]$. Let $r_0(w)$ be defined as
\begin{equation*}
r_0(w)=E[Y(1-D)\mid Z=1, w]-E[Y(1-D)\mid Z=0, w].
\end{equation*}

By using variation in $r_0(w)$ with respect to $w$, direct effect  $\rho_0$ and $E[Y_{0, 0}\mid \CP]$ can be identified as follows:
\begin{equation*}
\begin{aligned}
&\rho_0=\frac{r_0(1)\Pr(\CP\mid 0)-r_0(0)\Pr(\CP\mid 1) }{\Pr(\NT\mid 1)\Pr(\CP\mid 0)-\Pr(\NT\mid 0)\Pr(\CP\mid 1) }, \\
&E[Y_{0, 0}\mid \CP]=-\frac{r_0(1)-\rho_0\Pr(\NT\mid 1) }{\Pr(\CP\mid 1) }.\\
\end{aligned}
\end{equation*}

Therefore, LATE is identified since the two direct effects $(\rho_1, \rho_0)$ and $(E[Y_{1, 1}\mid \CP], E[Y_{0, 0}\mid \CP] )$ are shown to be identified. Plugging into the expression of $(E[Y_{1, 1}\mid \CP], E[Y_{0, 0}\mid \CP] )$ into the formula of LATE leads to following expression:
\begin{equation*}
\begin{aligned}
\LATE&=E[Y_{1, 1}-Y_{0, 0}\mid \CP]-\rho_1\Pr(Z=0)-\rho_0\Pr(Z=1) \\
&=\frac{r_1(1)+r_0(1)}{\Pr(\CP\mid 1)}-\rho_1 \left(\frac{\Pr(\AT \mid 1)}{\Pr(\CP\mid 1)} +\Pr(Z=0) \right)-\rho_0\left( \frac{\Pr(\NT \mid 1)}{\Pr(\CP\mid 1)}+ \Pr(Z=1) \right) \\
&\equiv \IV_1-\rho_1 w_1^{\rho}-\rho_0 w^{\rho}_0,
\end{aligned}
\end{equation*}
where $\IV_1=\frac{E[Y\mid Z=1, W=1]-E[Y\mid Z=0, W=1]}{E[D\mid Z=1, W=1]-E[D\mid Z=0, W=1]}$, $w_1^{\rho}= \frac{\Pr(\AT \mid 1)}{\Pr(\CP\mid 1)} +\Pr(Z=0)$, and $w_0^{\rho}= \frac{\Pr(\NT \mid 1)}{\Pr(\CP\mid 1)}+ \Pr(Z=1)$. Therefore, LATE is identified as all terms in the above expression are shown to be identified.
%
%

\end{proof}

\subsection{Asymptotic Properties of $\widehat{\LATE}$ and $(\hat{\rho}_1, \hat{\rho}_0)$} \label{asymptotic}

This section derives the asymptotic properties of estimator $\widehat{\LATE}$ for the local average treatment effect and the two estimators $(\hat{\rho}_1, \hat{\rho}_0)$ for the two direct effects. Suppose that the variances of all variables $(Y, D, Z, W)$ are finite, and the variances of the product of any two, three, and four variables are finite. 

As shown in Section \ref{sec:esti} , the expression for estimator $\widehat{\LATE}$ is given as follows:
\begin{equation*}
\widehat{\LATE}=\widehat{\IV}_1-\hat{\rho}_1 \hat{w}^{\rho}_1-\hat{\rho}_0 \hat{w}^{\rho}_0.
\end{equation*}

The analysis proceeds by deriving the asymptotic properties of the above terms $(\widehat{\IV}_1, \hat{w}^{\rho}_1, \hat{w}^{\rho}_0, \hat{\rho}_1, \hat{\rho}_0)$ respectively, then the asymptotic property of $\widehat{\LATE}$ can be derived accordingly.

 Let $\mu_Y$ denote the expectation of the random variable $Y$. Let $\phi_i^{Y}=Y_i-\mu_Y$ for any random variable $Y$, let $\phi_i^{YZ}=Y_iZ_i-E[YZ]$ for any two random variables $(Y, Z)$, and let $\phi_i^{YZW}=Y_iZ_iW_i-E[YZW]$ for any three random variables $(Y, Z, W)$.

I first look at estimator $\widehat{\IV}_1$. The numerator of estimator $\widehat{\IV}_1$ is given as
\begin{equation*}
\begin{aligned}
&\frac{1}{N}\sum_i(Y_iZ_iW_i) \bar{W}-\frac{1}{N}\sum_i(Y_iW_i)\frac{1}{N}\sum_i(Z_iW_i)-(E[YZW]E[W]-E[YW]E[ZW])\\
=&\frac{1}{N}\sum_i  (E[YZW]\phi_i^{W}+\mu_W\phi_i^{YZW}-E[YW]\phi_i^{ZW}-E[ZW]\phi_i^{YW})+O_p\left(\frac{1}{N} \right).
\end{aligned}
\end{equation*}

The last equality holds by applying Taylor expansion and the fact that $\frac{1}{N}\sum_i(Y_iZ_iW_i)-E[YZW]=O_p(1/\sqrt{N})$, $\bar{W}-\mu_W=O_p(1/\sqrt{N})$, $\frac{1}{N}\sum_i(Y_iW_i)-E[YW]=O_p(1/\sqrt{N})$, and $ \frac{1}{N}\sum_i(Z_iW_i)-E[ZW]=O_p(1/\sqrt{N})$.

Let $\phi_i^{\IV_N}$ denote the influence function of the numerator term $\frac{1}{N}\sum_i(Y_iZ_iW_i) \bar{W}-\frac{1}{N}\sum_i(Y_iW_i)\frac{1}{N}\sum_i(Z_iW_i)$, defined as
\begin{equation*}
\phi_i^{\IV_N}=E[YZW]\phi_i^{W}+\mu_W\phi_i^{YZW}-E[YW]\phi_i^{ZW}-E[ZW]\phi_i^{YW}.
\end{equation*}

Similarly, the influence function of the denominator term $\frac{1}{N}\sum_i(D_iZ_iW_i) \bar{W}-\frac{1}{N}\sum_i(D_iW_i)\frac{1}{N}\sum_i(Z_iW_i)$ is derived as 
\begin{equation*}
\phi_i^{\IV_D}=E[DZW]\phi_i^{W}+\mu_W\phi_i^{DZW}-E[DW]\phi_i^{ZW}-E[ZW]\phi_i^{DW}.
\end{equation*}

Then by applying Taylor expansion, the asymptotic property of $\widehat{\IV}_1$  is derived as follows:
\begin{equation*}
\begin{aligned}
\widehat{\IV}_1-\IV_1&=\frac{1}{N}\sum_i \left\{  \frac{\phi_i^{\IV_N}-\IV_1 \phi_i^{\IV_D} }{E[YZW]E[W]-E[YW]E[ZW]} \right\}+O_p\left(\frac{1}{N} \right) \\
&\equiv \frac{1}{N}\sum_i \phi_i^{\IV_1}+O_p\left(\frac{1}{N} \right).
\end{aligned}
\end{equation*}

The approach for deriving influence functions of the two weights and direct effects is similar. I focus on the properties of $\hat{w}^{\rho}_1$ and $\hat{\rho}_1$, and the analysis applies to $(\hat{w}^{\rho}_0, \hat{\rho}_0)$.  Estimator $\hat{w}_1^{\rho}$ is given as
\begin{equation*}
\begin{aligned}
\hat{w}^{\rho}_1=\frac{\widehat{\Pr}(\AT\mid 1)} {\widehat{\Pr}(\CP\mid 1)}+\widehat{\Pr}(Z=0).
\end{aligned}
\end{equation*}

To derive the asymptotic property of $\hat{w}^{\rho}_1$, I need to show the asymptotic property of the conditional probability of the two subgroups. 
Let $\phi_i^{\AT1}$, $\phi_i^{\AT0}$, $\phi_i^{\CP1}$, $\phi_i^{\CP0}$ denote the influence function for the four conditional probabilities $\widehat{\Pr}(\AT\mid  1)$, $\widehat{\Pr}(\AT\mid 0)$, $\widehat{\Pr}(\CP\mid 1)$, $\widehat{\Pr}(\CP\mid 0)$ respectively, which are derived as follows:
\begin{equation*}
\begin{aligned}
\phi_i^{\AT1}&=\frac{\phi_i^{D\tilde{Z}W}} {E[\tilde{Z}W]}- \frac{E[D\tilde{Z}W]\phi_i^{\tilde{Z}W}}{(E[\tilde{Z}W])^2}, \qquad 
\phi_i^{\AT0}=\frac{\phi_i^{D\tilde{Z}\tilde{W}}} {E[\tilde{Z}\tilde{W}]}- \frac{E[D\tilde{Z}\tilde{W}]\phi_i^{\tilde{Z}\tilde{W}}}{(E[\tilde{Z}\tilde{W}])^2}, \\
\phi_i^{\CP1}&=\frac{\phi_i^{DZW}}{E[ZW]}-\frac{E[DZW]\phi_i^{ZW}}{(E[ZW])^2}-\left\{ \frac{\phi_i^{D\tilde{Z}W}}{E[\tilde{Z}W]}-\frac{E[D\tilde{Z}W]\phi_i^{\tilde{Z}W}}{(E[\tilde{Z}W])^2} \right\}, \\
\phi_i^{\CP0}&=\frac{\phi_i^{DZ\tilde{W}}}{E[Z\tilde{W}]}-\frac{E[DZ\tilde{W}]\phi_i^{Z\tilde{W}}}{(E[Z\tilde{W}])^2}-\left\{ \frac{\phi_i^{D\tilde{Z}\tilde{W}}}{E[\tilde{Z}\tilde{W}]}-\frac{E[D\tilde{Z}\tilde{W}]\phi_i^{\tilde{Z}\tilde{W}}}{(E[\tilde{Z}\tilde{W}])^2} \right\},
\end{aligned}
\end{equation*}
where $\tilde{Z}=1-Z$ and $\tilde{W}=1-W$.

Then the influence function for $\hat{w}_1^{\rho}$ is derived as follows by applying Taylor expansion:
\begin{equation*}
\begin{aligned}
\hat{w}^{\rho}_{1}-w^{\rho}_1&=\frac{1}{N}\sum_i\left\{ \frac{ \phi_i^{\AT1}}{\Pr(\CP\mid 1)}-\frac{\Pr(\AT\mid 1)\phi_i^{\CP1}}{(\Pr(\CP\mid 1))^2}-\phi_i^Z \right\}+ O_p\left(\frac{1}{N}\right)  \\
&\equiv\frac{1}{N}\sum_i \phi_i^{w^{\rho}_1}+O_p\left(\frac{1}{N}\right).
\end{aligned}
\end{equation*}

Now we only need to derive the influence function of estimator $\hat{\rho}_1$, given as
\begin{equation*}
\hat{\rho}_1=\frac{\hat{r}_1(1)\widehat{\Pr}(\CP\mid 0)-\hat{r}_1(0)\widehat{\Pr}(\CP\mid 1) }{\widehat{\Pr}(\AT\mid 1)\widehat{\Pr}(\CP\mid 0)-\widehat{\Pr}(\AT\mid 0)\widehat{\Pr}(\CP\mid 1) }. 
\end{equation*}


Let $\phi_i^{d1}$ denote the influence function for the denominator of $\hat{\rho}_1$, derived as
\begin{equation*}
\phi_i^{d1}=\Pr(\AT\mid 1)\phi_i^{\CP0}+\Pr(\CP\mid 0) \phi_i^{\AT1}-\Pr(\AT\mid 0)\phi_i^{\CP1}-\Pr(\CP\mid 1)\phi_i^{\AT0}.
\end{equation*}

 I look at the numerator of estimator $\hat{\rho}_1$. Let $\phi_i^{YDZW}=Y_iD_iZ_iW_i-E[YDZW]$ for any random variables $(Y, D, Z, W)$.  According to the definition of $\hat{r}_1(1), \hat{r}_1(0)$, their influence functions $\phi_i^{r_{11}}$ and $\phi_i^{r_{10}}$ are shown as
\begin{equation*}
\begin{aligned}
\phi_i^{r_{11}}&=\frac{\phi_i^{YDZW}}{E[ZW]}-\frac{E[YDZW]\phi_i^{ZW}}{(E[ZW])^2}-\left\{\frac{\phi_i^{YD\tilde{Z}W}}{E[\tilde{Z}W]}-\frac{E[YD\tilde{Z}W]\phi_i^{\tilde{Z}W}}{(E[\tilde{Z}W])^2} \right\}, \\
\phi_i^{r_{10}}&=\frac{\phi_i^{YDZ\tilde{W} }}{E[Z\tilde{W} ]}-\frac{E[YDZ\tilde{W} ]\phi_i^{Z\tilde{W} }}{(E[Z\tilde{W} ])^2}-\left\{ \frac{\phi_i^{YD\tilde{Z}\tilde{W} }}{E[\tilde{Z}\tilde{W} ]}-\frac{E[YD\tilde{Z}\tilde{W} ]\phi_i^{\tilde{Z}\tilde{W} }}{(E[\tilde{Z}\tilde{W} ])^2} \right\}.
\end{aligned}
\end{equation*}

Let $\phi_i^{n1}$ denote the influence function of the numerator of $\hat{\rho}_1$, derived as
\begin{equation*}
\phi_i^{n1}=r_1(1)\phi_i^{\CP0}+\Pr(\CP\mid 0)\phi_i^{r_{11}}-r_1(0)\phi_i^{\CP1}-\Pr(\CP\mid 1)\phi_i^{r_{10}}.
\end{equation*}

Now we are ready to derive the influence function for estimator $\hat{\rho}_1$. Let $\phi_i^{\rho_1}$ denote the influence function for estimator $\hat{\rho}_1$, derived as follows:
\begin{equation*}
\phi_i^{\rho_1}=\frac{ \phi_i^{n1}-\rho_1 \phi_i^{d1} }{ \Pr(\AT\mid 1)\Pr(\CP\mid 0)-\Pr(\AT\mid 0)\Pr(\CP\mid 1) }.
\end{equation*}

Therefore, the asymptotic property of estimator $\hat{\rho}_1$ is derived as
\begin{equation*}
\sqrt{N}(\hat{\rho}_1-\rho_1)\rightarrow \mathcal{N}(0, \Var(\phi_i^{\rho_1} ) ).
\end{equation*}

The analysis for the estimators $\hat{\rho}_0$ and $\hat{w}^{\rho}_0$ can be shown similarly, so it is omitted here. Let $\phi_i^{\rho_0}$ and $\phi_i^{w^{\rho}_0}$ denote the influence function for $\hat{\rho}_0$ and $\hat{w}^{\rho}_0$ respectively. Then the influence function $\phi_i^{\LATE}$ for estimator $\widehat{\LATE}$ is expressed as
\begin{equation*}
\phi_i^{\LATE}=\phi_i^{\IV_1}-\rho_1\phi_i^{w^{\rho}_1}-w^{\rho}_1\phi_i^{\rho_1}-\rho_0\phi_{i}^{w^{\rho}_0}-w^{\rho}_0\phi_i^{\rho_0}.
\end{equation*}

The asymptotic property for $\widehat{\LATE}$ is shown as
\begin{equation*}
\sqrt{N}(\widehat{\LATE}-\LATE) \rightarrow \mathcal{N}(0, \Var(\phi_i^{\LATE})).
\end{equation*}

\subsection{Proof of Proposition \ref{prop:exte}} \label{proof:exte}

\begin{proof}

Following the proof in \ref{proof:thm1}, LATE can be divided into two groups $Z=1$ and $Z=0$ as follows:
\begin{equation*}
\begin{aligned}
\LATE=&E[Y_1-Y_0\mid \CP]\\
=&E[Y_{1, 1}-Y_{0, 1}\mid \CP, Z=1]\Pr(Z=1)+E[Y_{1, 0}-Y_{0, 0}\mid \CP, Z=0]\Pr(Z=0)\\
=&E[Y_{1, 1}-Y_{0, 1}\mid \CP]\Pr(Z=1)+E[Y_{1, 0}-Y_{0, 0}\mid \CP]\Pr(Z=0).
\end{aligned}
\end{equation*}

Using $\rho_{\CP, d}=E[Y_{d, 1}-Y_{d, 0}\mid \CP]$ to substitute $E[Y_{0, 1} \mid \CP]$ and $E[Y_{1, 0} \mid \CP]$ has the following implication:
\begin{equation*}
\LATE=E[Y_{1, 1}-Y_{0, 0}\mid \CP]-\rho_{\CP, 1}\Pr(Z=0)-\rho_{\CP, 0} \Pr(Z=1).
\end{equation*}

As shown in \ref{proof:thm1}, using variation in $r_1(w)$ with respect to $w$ can identify $\rho_{\AT, 1}$ and $E[Y_{1,1}\mid \CP]$ as follows:
\begin{equation*}
\begin{aligned}
&\rho_{\AT, 1}=\frac{r_1(1)\Pr(\CP\mid 0)-r_1(0)\Pr(\CP\mid 1) }{\Pr(\AT\mid 1)\Pr(\CP\mid 0)-\Pr(\AT\mid 0)\Pr(\CP\mid 1) }, \\
&E[Y_{1,1}\mid \CP]=\frac{r_1(1)-\rho_1\Pr(\AT\mid 1) }{\Pr(\CP\mid 1 ) }. \\
\end{aligned}
\end{equation*}

Similarly, the direct effect  $\rho_{\NT, 0}$ and $E[Y_{0 ,0}\mid \CP]$ are identified by exploiting variation in $r_0(w)$:
\begin{equation*}
\begin{aligned}
&\rho_{\NT, 0}=\frac{r_0(1)\Pr(\CP\mid 0)-r_0(0)\Pr(\CP\mid 1) }{\Pr(\NT\mid 1)\Pr(\CP\mid 0)-\Pr(\NT\mid 0)\Pr(\CP\mid 1) }, \\
&E[Y_{0, 0}\mid \CP]=-\frac{r_0(1)-\rho_0\Pr(\NT\mid 1) }{\Pr(\CP\mid 1) }.\\
\end{aligned}
\end{equation*}

Under Assumption \ref{ass:dir} about the difference in direct effects across subgroups, $\rho_{\CP, d}$ for $d\in\{0, 1\}$ can be bounded as
\begin{equation*}
\begin{aligned}
\rho_{\AT, 1}-k_1\leq \rho_{\CP, 1}\leq \rho_{\AT, 1}+k_1, \\
\rho_{\NT, 0}-k_0\leq \rho_{\CP, 0}\leq \rho_{\NT, 0}+k_0. \\
\end{aligned}
\end{equation*}

Therefore, the bounds on $\LATE$ is established as follows:
\begin{equation*}
\begin{aligned}
\LATE \geq E[Y_{1, 1}-Y_{0, 0}\mid \CP]-(\rho_{\AT, 1}+k_1)\Pr(Z=0)-(\rho_{\NT, 0}+k_0) \Pr(Z=1), \\
\LATE \leq E[Y_{1, 1}-Y_{0, 0}\mid \CP]-(\rho_{\AT, 1}-k_1)\Pr(Z=0)-(\rho_{\NT, 0}-k_0) \Pr(Z=1).
\end{aligned}
\end{equation*}

Plugging into the formulas for $(\rho_{\AT, 1}, \rho_{\NT, 0})$ and $(E[Y_{1, 1}\mid \CP], E[Y_{0 ,0}\mid \CP])$ lead to the results in Proposition \ref{prop:exte}. The lower bound is achieved when $\rho_{CP, 1}=\rho_{\AT, 1}+k_1$ and $\rho_{CP, 0}=\rho_{\NT, 0}+k_0$, and the upper bound is achieved when $\rho_{CP, 1}=\rho_{\AT, 1}-k_1$ and $\rho_{CP, 0}=\rho_{\NT, 0}-k_0$. Therefore, the bounds are sharp.

\end{proof}

\subsection{More Simulation Results} \label{simu}

This section presents the simulation results of the two estimators $(\hat{\theta}_{zw}, \hat{\theta}_z)$ when the size of the subgroups changes. Consider that the potential treatments $(D_1, D_0)$ is given as follows:
\begin{equation*}
\begin{aligned}
D_1&=\mathbbm{1}\{\epsilon\leq 1   \}, \\
D_0&=\mathbbm{1}\{\epsilon\leq k   \}.
\end{aligned}
\end{equation*}

The value of $k$ determines the size of the three subgroups, and the size of compliers $\Pr(k<\epsilon<1)$ decreases when the value of $k$ increases. 
I consider three specifications of $k$: $k\in \{-0.25, 0, 0.25\}$. 

Table \ref{table3} and \ref{table4} show the results of the two estimators $(\hat{\theta}_{zw}, \hat{\theta}_z)$ under different values of $k$ and different values of direct effects $\rho$ with the sample size $N=1000$ and $N=4000$, respectively. The two estimators  $(\hat{\theta}_{zw}, \hat{\theta}_z)$ both perform better when the size of compliers increases, while the estimator $\hat{\theta}_{zw}$ uniformly performs better than the estimator $\hat{\theta}_z$ with nonzero direct effects regardless of the size of compliers. Therefore, the robustness feature of the estimator $\hat{\theta}_{zw}$ to nonzero direct effects still holds under different probabilities of subgroups.

\begin{table}[!htbp]
\centering
\caption{Performance Comparisons of $\hat{\theta}_{zw}$ and $\hat{\theta}_z$ $(N=1000)$  }
\label{table3}
\begin{tabular}{c |cccc|cccc}
\hline
\hline
\multirow{2}{*}{$\rho$}&\multicolumn{4}{c|}{$\hat{\theta}_{zw}$}  &  \multicolumn{4}{c}{$\hat{\theta}_z$}  \\
\cline{2-9} 
  & Bias & SD & rMSE & MAD  & Bias & SD  & rMSE & MAD \\
\hline
\multicolumn{9}{c}{$k=-0.25$} \\
\hline
 $0$ \ & 0.016 & 0.459 & 0.459 & 0.362  & 0.004 &0.142 &0.142 &0.113  \\ [1.0ex]       
$0.5$ \ & 0.016 & 0.459 & 0.459 & 0.362 &1.145 & 0.173 & 1.158 &1.145  \\ [1.0ex]                                                                                                                                                                                                                                                                                                                                                                                                                                                                                                                                                                                                                                                                     $1$ \  & 0.016 & 0.459 & 0.459 & 0.362  & 2.285 & 0.225 & 2.296 & 2.285 \\ [1.0ex]  
$-1$ \  & 0.016 & 0.459 & 0.459 & 0.362  & -2.277 & 0.178 &2.284 &2.277  \\ [1.0ex] 
\hline
\multicolumn{9}{c}{$k=0$} \\
\hline
 $0$ \ & 0.025 & 0.504 & 0.505 & 0.396 & 0.006 &0.183 &0.184 &0.146  \\ [1.0ex]       
$0.5$ \ & 0.025 & 0.504 & 0.505 & 0.396 &1.481 & 0.245 & 1.501 &1.481  \\ [1.0ex]                                                                                                                                                                                                                                                                                                                                                                                                                                                                                                                                                                                                                                                                     $1$ \  & 0.025 & 0.504 & 0.505 & 0.396  & 2.956 & 0.342 & 2.976 & 2.956 \\ [1.0ex]  
$-1$ \  & 0.025 & 0.504 & 0.505 & 0.396  & -2.944 & 0.269 &2.956 &2.944  \\ [1.0ex] 
\hline
\multicolumn{9}{c}{$k=0.25$} \\
\hline
 $0$ \ &0.057 &0.915 &0.917 & 0.508 & 0.010 &0.262 &0.262 &0.207  \\ [1.0ex] 
 $0.5$ \  &0.057 &0.915 &0.917 & 0.508 & 2.100 &0.403 & 2.139 & 2.100   \\ [1.0ex] 
 $1$ \  &0.057 &0.915 &0.917 & 0.508 &  4.190 & 0.615 &4.235 & 4.190     \\ [1.0ex] 
 $-1$ \  &0.057 &0.915 &0.917 & 0.508 & -4.169 &0.498 &4.199 &4.169     \\   [1.0ex] 

\hline
\end{tabular}
\end{table}

\begin{table}[!htbp]
\centering
\caption{Performance Comparisons of $\hat{\theta}_{zw}$ and $\hat{\theta}_z$ $(N=4000)$  }
\label{table4}
\begin{tabular}{c |cccc|cccc}
\hline
\hline
\multirow{2}{*}{$\rho$}&\multicolumn{4}{c|}{$\hat{\theta}_{zw}$}  &  \multicolumn{4}{c}{$\hat{\theta}_z$}  \\
\cline{2-9} 
  & Bias & SD & rMSE & MAD  & Bias & SD  & rMSE & MAD \\
\hline
\multicolumn{9}{c}{$k=-0.25$} \\
\hline
 $0$ \  & 0.007 & 0.214 & 0.214 & 0.170  & -0.000 &0.072 &0.072 &0.057  \\ [1.0ex]       
$0.5$ \ & 0.007 & 0.214 & 0.214 & 0.170 &1.136 & 0.086 & 1.139 &1.136  \\ [1.0ex]                                                                                                                                                                                                                                                                                                                                                                                                                                                                                                                                                                                                                                                                     $1$ \  & 0.007 & 0.214 & 0.214 & 0.170   & 2.273 & 0.110 & 2.275 & 2.273 \\ [1.0ex]  
$-1$ \  & 0.007 & 0.214 & 0.214 & 0.170  & -2.274 & 0.089 &2.275 &2.274  \\ [1.0ex] 
\hline
\multicolumn{9}{c}{$k=0$} \\
\hline
 $0$ \ &0.009 &0.230 &0.230 & 0.183 &-0.001 &0.092 &0.092 &0.073  \\ [1.0ex] 
 $0.5$ \  &0.009 &0.230 &0.230 & 0.183 & 1.466 &0.119 & 1.470 & 1.466   \\ [1.0ex] 
 $1$ \  &0.009 &0.230 &0.230 & 0.183&  2.932 & 0.165 &2.936 &2.932      \\ [1.0ex] 
$-1$ \  &0.009 &0.230 &0.230 & 0.183 & -2.933 &0.133 &2.936 &2.933     \\   [1.0ex] 
\hline
\multicolumn{9}{c}{$k=0.25$} \\
\hline
 $0$ \   &0.013 &0.273 &0.273 & 0.217 & -0.002 &0.130  &0.130 &0.104  \\ [1.0ex] 
 $0.5$ \  &0.013 &0.273 &0.273 & 0.217 & 2.064 &0.193 & 2.073 & 2.064   \\ [1.0ex] 
 $1$ \  &0.013 &0.273 &0.273 & 0.217 &  4.129 & 0.291 &4.140 & 4.129     \\ [1.0ex] 
 $-1$ \  &0.013 &0.273 &0.273 & 0.217 & -4.133 &0.242 &4.140 &4.133     \\   [1.0ex] 

\hline
\end{tabular}
\end{table}

\end{document}